# Binding Pathway of Opiates to μ-Opioid Receptors Revealed by Unsupervised Machine Learning

*Amir Barati Farimani[§], Evan N. Feinberg[§], Vijay S. Pande[1]*

Department of Chemistry, Stanford University, Stanford, California 94305

[§] The authors equally contributed to this work

**Abstract**

Many important analgesics relieve pain by binding to the μ-Opioid Receptor (μOR), which makes the μOR among the most clinically relevant proteins of the G Protein Coupled Receptor (GPCR) family. Despite previous studies on the activation pathways of the GPCRs, the mechanism of opiate binding and the selectivity of μOR are largely unknown. We performed extensive molecular dynamics (MD) simulation and analysis to find the selective allosteric binding sites of the μOR and the path opiates take to bind to the orthosteric site. In this study, we predicted that the allosteric site is responsible for the attraction and selection of opiates. Using Markov state models and machine learning, we traced the pathway of opiates in binding to the orthosteric site, the main binding pocket. Our results have important implications in designing novel analgesics.

**Introduction**

The most powerful analgesic and addictive properties of opiates are mediated by the μ-Opioid Receptor (μOR).[1,2] Since this receptor is primarily responsible for the effects of opium, the μOR

---

[1] Corresponding Author, e-mail: pande@stanford.edu, web: https://pande.stanford.edu/

is one of the oldest drug targets for the discovery of analgesics.[3] µOR activation results in signaling through the heterotrimeric G protein $G_i$, resulting in analgesia and sedation. Activity studies of µOR have revealed that subtle changes in ligand structure can make the difference between an agonist and an antagonist, so there is a general philosophy within the GPCR drug community that distinct pharmacophores are responsible for efficacy (message) or selectivity (address).[4, 5] An efficient opioid (medicinally perfect) would relieve pain potently without side effects such as harmful respiratory effects or constipation, show sustained efficacy in chronic treatments, and not cause addiction.[6, 7] Design of such an ideal opioid requires a fundamental and deep understanding of drug binding, dynamics, and receptor activation mechanism.[8, 9, 10]

In spite of significant studies in the past few years on the conformational changes triggered due to drug binding,[11] the origin of the selectivity of certain receptors to specific drugs is largely unknown.[9, 12-14] Does this selectivity originate from the binding pocket or are there other significant selectivity sites that prescreen the drug? Following this line of thought, it is important to answer the question regarding the binding dynamics, as well as the path that the drug travels from the extracellular environment to the binding pocket. It is still unknown how drugs diffuse through a highly tortuous cavity to arrive at the binding pocket, also called the orthosteric site.

To address the above questions, we used molecular dynamics (MD) simulations with the state of the art post-processing software MSMBuilder[15] and machine learning (ML) algorithm Time-Structure Based Independent Component Analysis (tICA).[16,17] First, we found an allosteric site responsible for the selectivity of µOR through monitoring the binding ability of different ligands. Second, we tried to understand the pathway of a ligand from the allosteric site (selectivity site) to the orthosteric site (main binding pocket). Since ligands need to assume specific and rare orientations to permeate through the tortuous and high-energy barrier cavity of the receptor, conventional long-trajectory MD simulations, with continuous trajectory lengths up to 2 μs, starting from allosteric site usually cannot show the full binding pathway; therefore, using Markov state models (MSMs) and tICA would be beneficial to unravel the binding dynamics. The identifications of the screening mechanism of μOR and the permeation dynamics of opiates have important implications in designing analgesics because they need to have a specific fingerprint and molecular architecture to be recognized by the allosteric screening.

**Methods**

We performed MD simulations with NAMD 2.[18] A typical simulation set up consisted of the receptor (inactive structure, PDB code: 4DKL), lipid bilayer, water, and ions (~98000 atoms; Supporting Information Figure S1). The protein structure was then aligned to its predicted membrane pose based on its entry in the Orientations of Proteins in Membranes Database.[19] A lipid bilayer (POPC) membrane was created (90 Å × 90 Å) using Visual Molecular Dynamics software.[18] Subsequently, the protein was inserted in the membrane and 4 ligands of the same type were placed randomly on the extracellular section of the membrane. For each ligand in (BU72, morphine, oxymorphone, IBNtxA, and carazolol), a set of 25 simulations were created (see Supporting Information, Figure S1). After placing the ligands in the simulation box, the entire system, including the receptor, membrane, and ligands, was solvated by a 25 Å thick slab of water on each side of membrane. We ran the simulation for 40 ns to equilibrate the system of lipid bilayer and protein. The ionic concentration of NaCl was 0.15 M. We used the CHARMM36 force field parameters for the protein, TIP3P water molecules, and ions. We parametrized the ligands using Forcefield Toolkit (FFTK).[20] All quantum mechanical (QM) calculations were carried out in the Gaussian software.[21] First, we optimized the geometry of the ligand and then computed the partial charges in presence of water to account for the solvent-ligand dipole effect. We broke the molecule into three parts for the torsion scan because electronic structure calculation for a ligand with 49-58 atoms is computationally expensive. We scanned the torsion angles with 5° iterations. The MP2/6-31 G* level of theory and basis set is used for parametrization, optimization, charge calculation, and torsion scan. The SHAKE algorithm was used to maintain the rigidity of the water molecules. Periodic boundary conditions were applied in all three directions. The cutoff distance for the Lennard-Jones (LJ) interactions was 15 Å. The long-range electrostatic interactions were computed using the Particle-Mesh-Ewald (PME) method. The time step was selected to be 2 fs. For each simulation, energy minimization was performed for 100,000 steps. The system was then equilibrated for 5 ns in the NPT ensemble at 1 atm pressure and 300 K temperature. NPT simulation ensures that the water concentration is equal to the bulk value of 1 g/cm$^3$. Temperature was maintained at 300 K by applying the Nosè-Hoover thermostat with a time constant of 0.1 ps. In total, 125 simulations were run, each of length 500 ns, producing an aggregate trajectory length of 72.5 μs for the allosteric affinity simulation data set.

The second set of simulations were performed to find the pathway between the allosteric and orthosteric sites. We do not have prior knowledge of the pathway, but we know the possible allosteric sites and the definite orthosteric pocket. While all the hypothetical paths from the allosteric site to the orthosteric site are spatially short and traverse a sterically constricted region, it is possible to generate different ligand orientations and positions along these hypothetical paths. We generated 420 orientations/positions between the allosteric and orthosteric sites. The algorithm for generation of different orientations/positions is as follows: 1. Starting from the allosteric site, rotate the ligand about x, y, and z axes, every 15°. 2. Move the ligand closer to orthosteric site by 1 Å each iteration and repeat step 1. Using this algorithm, we were able to account for many possible initial configurations of the ligand along the path.

Analysis was driven by the Conformation software package,[22] a custom Python code for analysis of MD simulations, that leverages MSMBuilder[23] and MDTraj.[24] The package was modified specifically to incorporate ligand-protein interaction features, in this case for a large GPCR MD dataset. All residue-ligand pairs within 6.6 Å measured by closest heavy atom distance in either crystal structure were selected. Then, for each of these approximate residue-ligand pairs, both the closest heavy atom distance and Cα distance were computed for each frame in each trajectory. Next, the Sparse tICA algorithm was applied to determine the reaction coordinates, or slowest collective degrees of freedom (up to the ten slowest in this case), of the protein and ligand. A k-means model was trained with k=1,000 clusters. Finally, an MSM was constructed with lag time 25 ns and prior counts $1 \times 10^{-5}$. The equilibrium state probabilities from the MSMs were used individually in each condition to generate the free energy surfaces projected onto the features and tICA coordinates (tICs).

**Results and Discussion**

1. **Characterization of the Key Motifs in attracting Drug from Solvent to the Receptor**

In the first set of simulations, we ran simulations for all the ligands (BU72, morphine, oxymorphone, IBNtxA, carazolol) to identify the allosteric site and the average time it takes for the ligands to stably bind to the allosteric site (we call this "time to bind"). We set the root-mean-square deviation (RMSD) and binding time criteria to <7.5 Å and >100 ns, respectively, and checked that the binding was within 5 Å of the extracellular lumen of μOR. The RMSD criteria of 7.5 Å was selected because the average RMSD of BU72 in all binding events is 7.43 Å. As mentioned in the method section, we performed 25 simulations for finding the allosteric binding sites of BU72 and other ligands and the 7.5 Å criteria is the average RMSD of BU72 center of mass (COM) while bound to allosteric site. (see the supporting binding movie). The histogram plot of time to bind for BU72 shows the lowest time to bind among all the tested molecules (Figures 1a, 1b, 1c, and 1d). Among 25 simulations for IBNtxA, only one binding event is observed (Figure 1d). It has been shown that IBNtxA has a high affinity to bind 6 transmembrane (6TM) splice variant µOR but significantly less so to the 7TM splice variant[25] and this agrees with our results (Figure 1d). We also simulated carazolol as a control to see if the allosteric site of µOR can bind to non-opioid ligands. Carazolol is a high affinity antagonist to the β-adrenergic receptor.[26] In 25 binding simulations with carazolol, we did not observe any binding to the µOR (Figure 1d). The order of binding events out of 25 samples for opioids is: BU72>oxymorphone>morphine>IBNtxA (Figure 1e). To compare the motion characteristics of ligand in solvent, allosteric, orthosteric inactive (PDB: 4DKL), and orthosteric active (PDB: 5CML), we computed the RMSD of BU72 in each case (Figure 1f). The average RMSDs of BU72 are 33.35 Å, 7.43 Å, 3.44 Å, and 2.12 Å in solvent, allosteric site, orthosteric (inactive), and orthosteric (active), respectively. The RMSD is computed for the COM of BU72 with respect to aligned µOR. The ligand's RMSD decreases gradually from solvent to allosteric site and then to the main binding pocket.

We also looked at the hydration of ligand during its binding, from solvent to the allosteric site (Figure 2a). We counted the waters within 5 Å of BU72 in each frame of the simulation. On average, ~40 water molecules encapsulated BU72 when it was in the bulk solvent. Upon binding, BU72 dehydrated and lost half of its water (in the allosteric site, BU72 was accompanied by ~20 water molecules); compared to when it was in the orthosteric (active) site, BU72 had four times more water molecules (Figure 2a). The higher hydration level of BU72 in the allosteric site was due to its exposure to the solvent in the extracellular part of the µOR.

The schematics and detailed interaction maps of BU72 in the allosteric and orthosteric sites are depicted in Figure 2b and Figure 2c. Figure 2b shows a completely different orientation of BU72 in the allosteric and orthosteric sites. There is a 17.86 Å of distance between the COM of BU72 in allosteric site and orthosteric site. Initially and before the stable binding of BU72 to allosteric binding site, the ligand is captured by certain motifs in the extracellular upper loop residues (GLU310$^{1.93}$, GLN212$^{2.60}$). The allosteric site is composed of residue groups (GLU310$^{1.93}$, PHE313$^{7.30}$, GLN314$^{7.31}$, THR315$^{7.32}$, THR311$^{7.28}$), (ARG211$^{9.54}$, GLN212$^{8.88}$, SER214$^{9.40}$, ASP216$^{2.87}$, THR218$^{ECL2}$), and (ASN127$^{2.63}$, TYR128$^{2.64}$, MET130$^{2.66}$, GLY131$^{ECL1}$), which are shown in Figure 2c.

## 2. Machine Learning Revealed the Pathway from Allosteric Site to Orthosteric Site

In long time trajectories of nearly 1 μs, we did not observe an event of allosteric to orthosteric transition of BU72. This implies that the transition timescale from the allosteric site to the orthosteric site may not be tractable with unbiased MD simulation alone. To characterize the transition pathway of the ligand from the allosteric site to the orthosteric site with an unknown timescale, we used an unsupervised machine learning algorithm – Sparse tICA[27] – in conjunction with MSMs.[15, 28, 29] tICA analyzes large datasets of trajectories in an unbiased way to find the slowest reaction coordinates sampled by the system.[16] Since such reaction coordinates are defined by linear combinations of features, the choice of featurization played a key role in this analysis. Custom code was written and incorporated into the open source MDTraj[24] package that automates the measurement of pairwise distances between individual heavy atoms on simulated ligands and protein residues with which it might interact. In addition to its role in interpretation, tICA also served as a dimensionality reduction technique for discretizing the protein into conformational states (k-means) that were then used to construct an MSM. The MSM of binding, which we include in the Supporting Information, enables biophysicists and medicinal chemists alike to query a compendium of partial trajectories that link the allosteric and orthosteric binding pockets of µOR.

The free energy landscape of orientation-position of BU72 shows a convoluted path between allosteric and orthosteric sites (Figure 3). The aggregate MD data is projected onto x and y-axes, tIC1 and tIC2, the first and second slowest degree of freedom, respectively, accessible to the ligand in simulation. While tIC1 represents the distance between allosteric and orthosteric sites, tIC2 represents the orientation of BU72 with respect to the protein aligned structure. The two valleys next to each other (at approximate positions (0.5, -0.5), (1.3, -1.1) in arbitrary units) are the allosteric sites from which BU72 starts to travel to the orthosteric site with the coordinates of (-0.2, -1) in Figure 3. Another metastable well is observed at (1.0, 2.0) which has the characteristic of the initial drug hooking site. The highest energetic barrier on the path is located at (-0.1, 0.8). The features (residues) that were involved in the hooking site and allosteric site are elaborated on Figure 2c and 2d.

Using transition path theory (TPT),[30, 31] we determined some of the most probable paths from allosteric to orthosteric sites (Figure 3). These paths originated from allosteric site 1 or site 2 to the orthosteric site and passed through the high-energy barriers on the free energy landscape. The network map of the connections between MSM clusters shows multiple possible paths connecting the orthosteric site to the allosteric sites (Figure 4). Between the two allosteric sites, site 2 is more populated than site 1 (Figure 4). Some of the paths (see Supporting Information for the paths' cluster numbers) suggest that BU72 starts at site 1, traverses through site 2, and then reaches the orthosteric site. The orientation of BU72 shows the rotation of the phenolic hydroxyl group toward the receptor's cytosol (Figure 5a). Snapshots taken from the MSM states on the paths show that the $N$-methyl group of BU72 gradually becomes perpendicular to membrane plane (Figure 5b).

Initially and at allosteric sites, the BU72 *N*-methyl is almost parallel to the membrane plane (Figure 5b). We also characterized the pose and the neighboring residues to BU72 at the highest energy clusters on the free energy plot (Figure 5c). We observed the bond stretch (on an average, 0.8 Å) between phenyl group and BU72 at this state (high-energy barrier state, Figure 5c), which we attribute to the strong cavity confinement. The unusual crystallographic state of BU72 was observed in the ref.[5] The important residues that were within 5 Å of BU72 and were selected by tICA are shown in Figure 5c. The residues in highly constricted path are LEU219$^{45.52}$, TRP318$^{7.35}$, ILE296$^{6.51}$, ILE322$^{7.38}$, ASP147$^{3.32}$, TYR146$^{3.33}$, TYR326$^{7.43}$, and TRP293$^{6.48}$ (Figure 5c).

**Conclusions**

In summary, we identified the fingerprint that defines the selectivity motif of µOR and is responsible for screening the drug and accepting the opiate scaffold. This motif is comprised of amino acids that are perched on top of the extracellular part of the receptor. We also showed the existence of allosteric sites that are responsible for stabilizing the ligand binding to the receptor and guiding the ligand to the main binding pocket at the orthosteric site. The allosteric sites of µOR are only selective to opioids. The path from allosteric to orthosteric site is a tortuous path with high-energy barriers. Using ML and MSMs, we mapped the free energy landscape of a ligand traversing from the allosteric site to the orthosteric site. We characterized the ligand's pose at different states along the path and demonstrated the rotation of the *N*-methyl group toward the polar cavity of the orthosteric site.

**Competing Financial Interests**

V.S.P. is a consultant and SAB member of Schrodinger, LLC and Globavir, sits on the Board of Directors of Apeel Inc, Freenome Inc, Omada Health, Patient Ping, Rigetti Computing, and is a General Partner at Andreessen Horowitz.

**Acknowledgments**

The authors express their appreciation Franklin Lee for insightful input while crafting this manuscript. We also would like to thank Kilian Cavalotti, Brian Roberts, Jimmy Wu, and Stephane Thiell for their indispensable computing support. We thank NIH training grant T32 GM08294 and NIH grant 1171245-309-PADPO for funding, as well as acknowledge the use of the Blue Waters Supercomputer. We thank The Blue Waters Graduate Fellowship for its support of Evan Feinberg during this research.

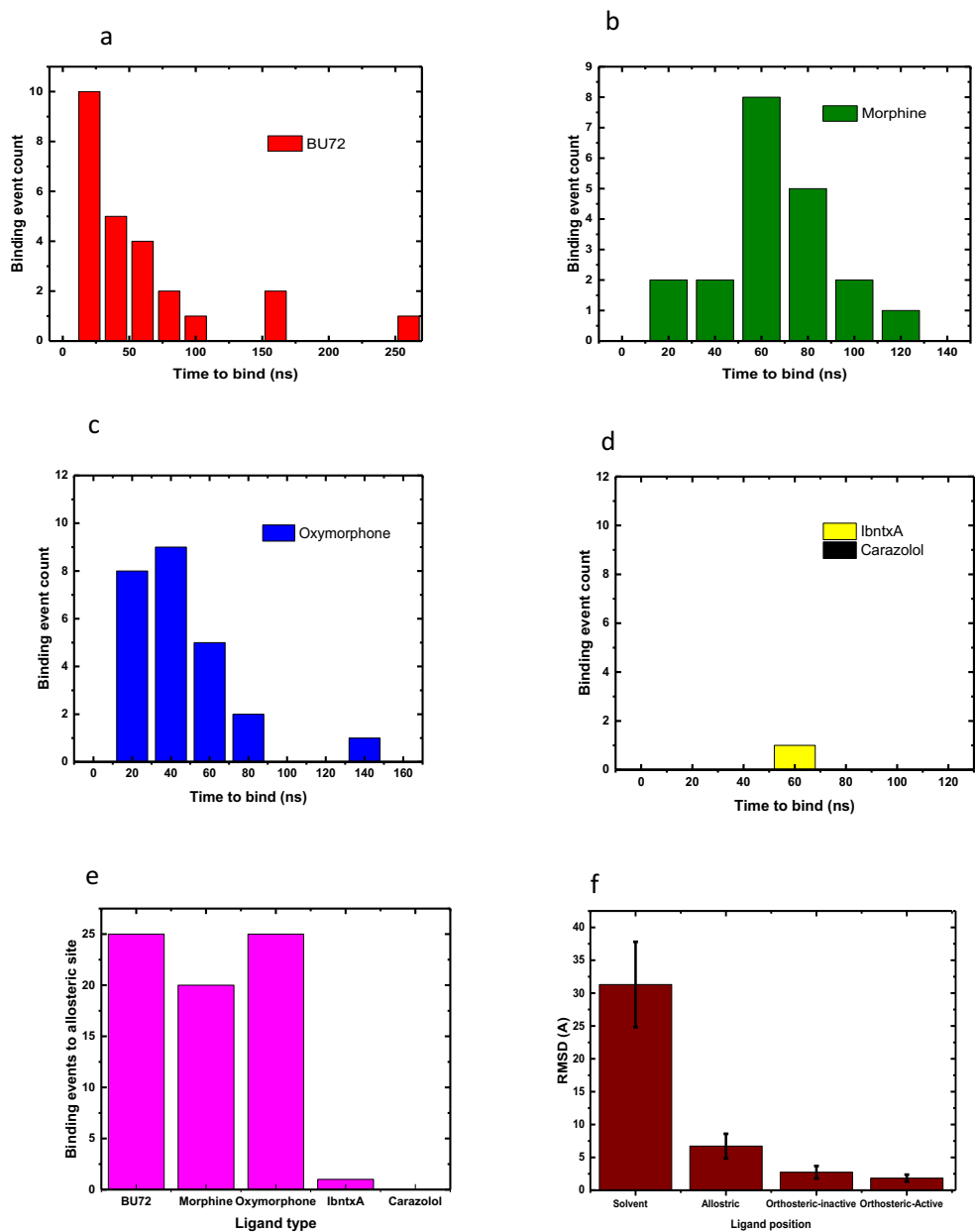

**Figure 1: a|** Binding time distribution from solvent to allosteric site of inactive µOR for BU72. **b|** Binding time from solvent to allosteric site of inactive µOR for morphine. **c|** Binding time from solvent to allosteric site of inactive µOR for oxymorphone. **d|** Binding time from solvent to allosteric site of inactive µOR for both IBNtxA and carazolol. **e|** Number of binding events to allosteric sites out of 25 simulations. **f|** Comparison of ligand RMSD in different locations with respect to the receptor.

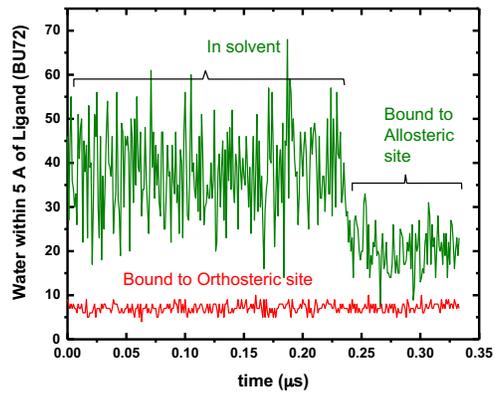

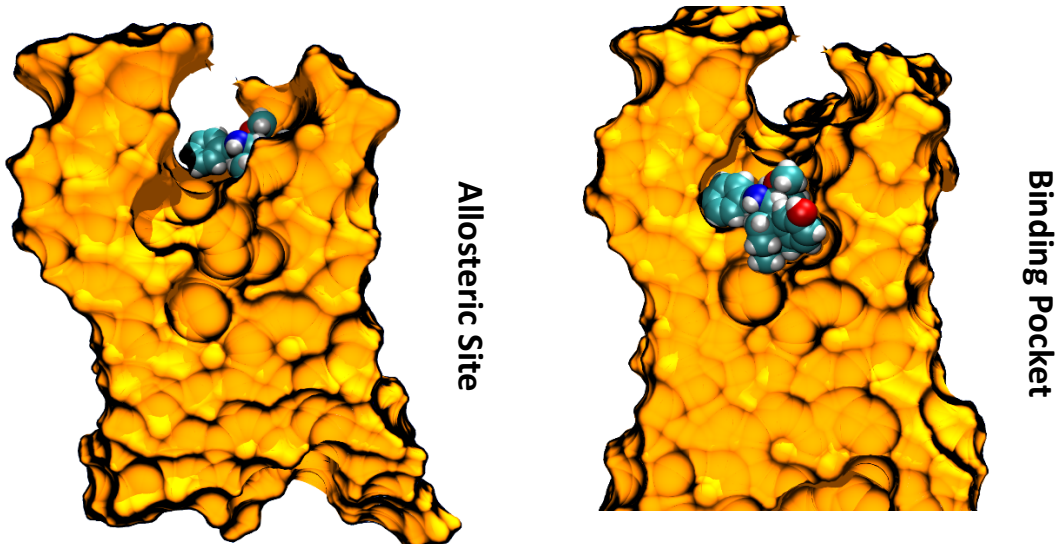

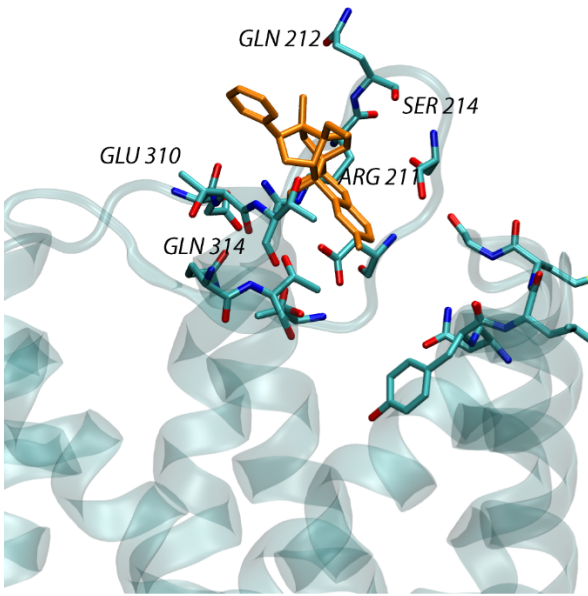
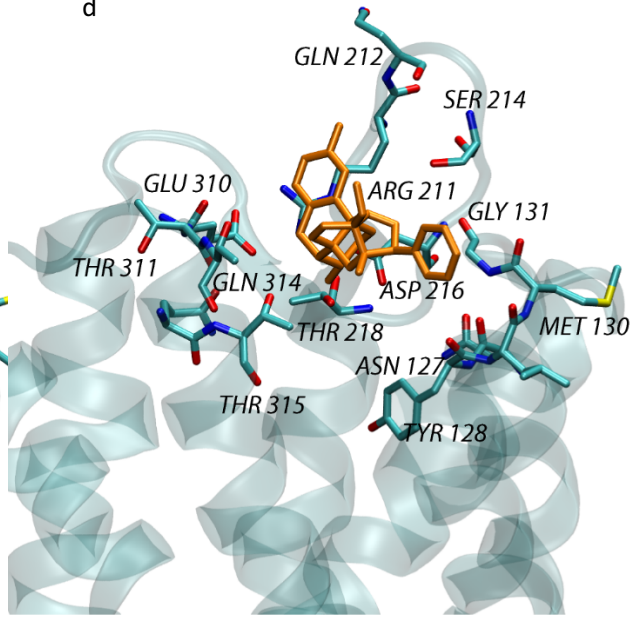

**Figure 2: a|** Comparison of BU72 hydration in solvent and in an allosteric site with hydration in the orthosteric site of μOR. **b|** Schematic comparison of the allosteric and orthosteric sites and the respective orientation and the distance between two sites. **c|** The residues involved in the initial "hooking mechanism" of the ligand. **d|** Allosteric binding pocket residues, the selectivity filter, and relative orientation of BU72 in the allosteric binding pocket.

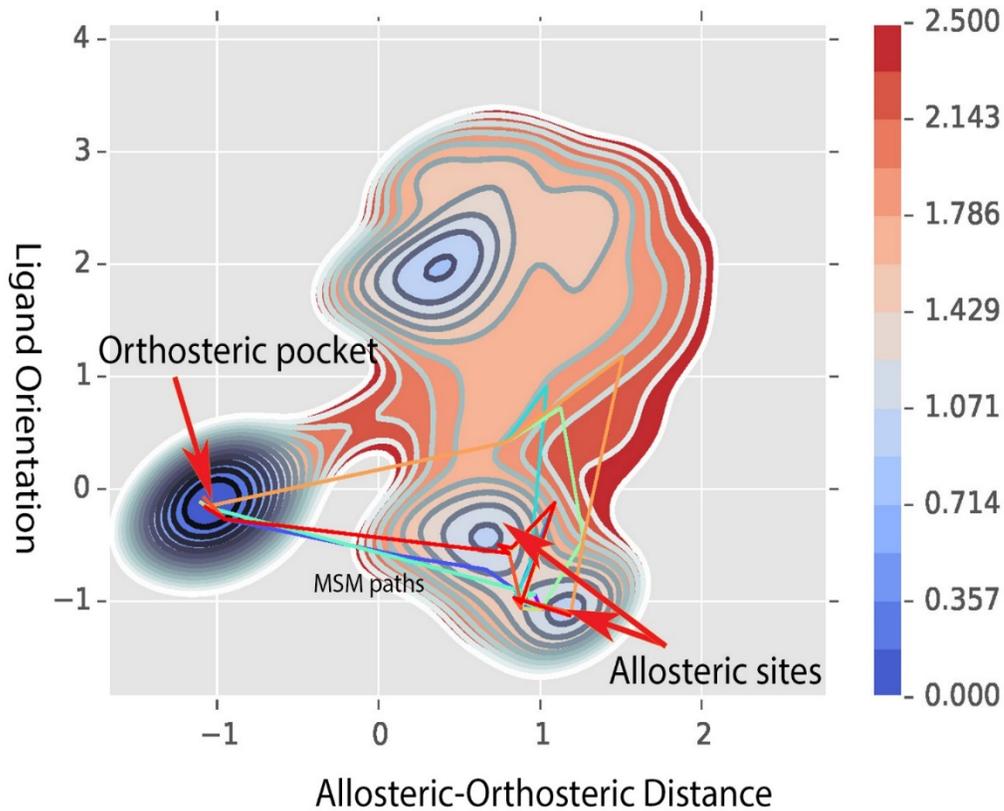

**Figure 3: a|** Free energy landscape of BU72 travelling for two allosteric sites to the orthosteric site. The x-axis represents the distance between the orthosteric and allosteric sites. The y-axis represents the ligand orientation with respect to fixed protein structure. The orange, light blue, blue, and red lines on the plot represent the most probable paths originated from TPT.

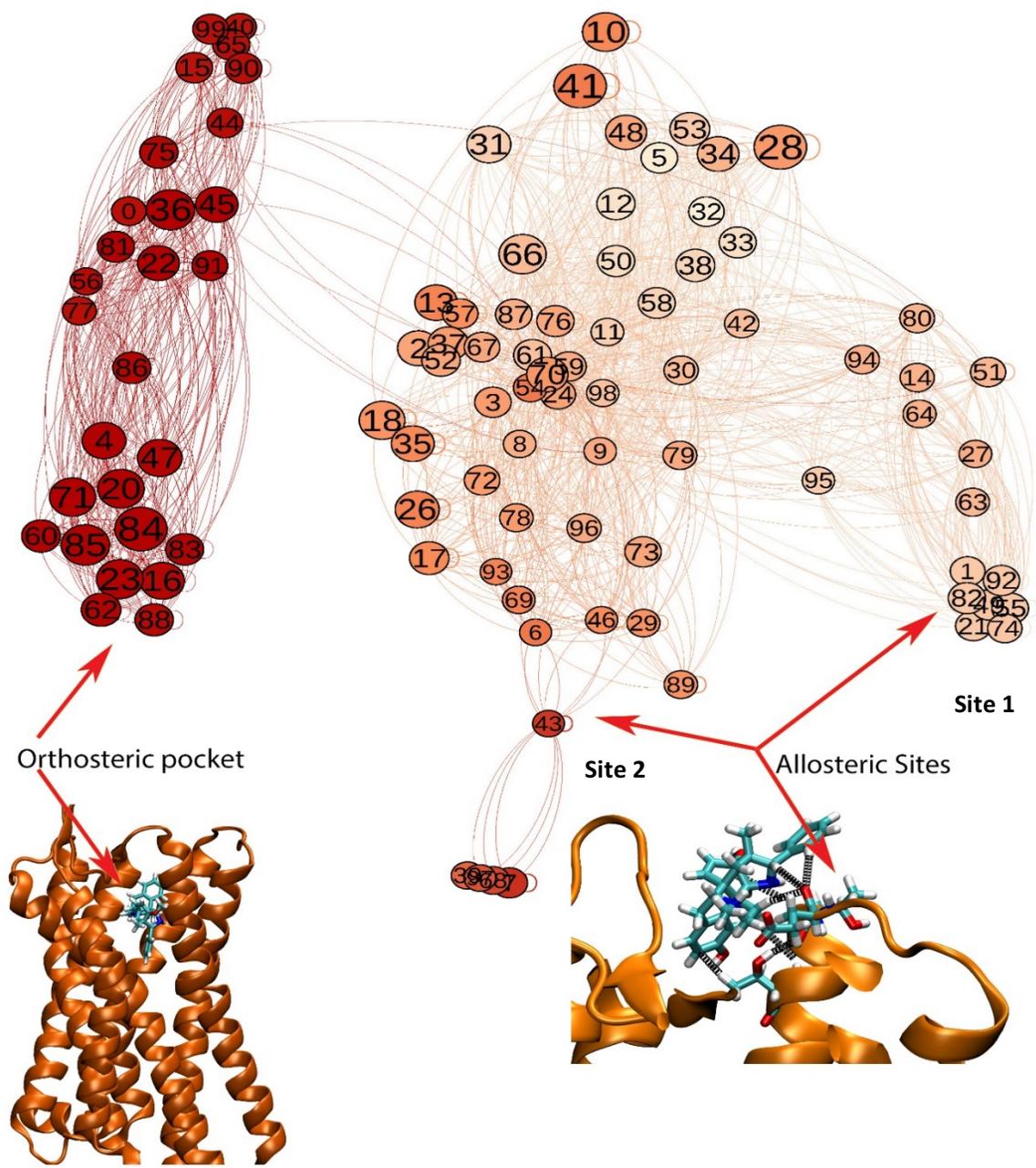

**Figure 4:** Network view of the BU72- μOR MSM states depicting allosteric-orthosteric transition. Each node is an MSM state and labeled based on the cluster number. The edges (connecting lines) are possible transition paths. The larger nodes represent more populated MSM clusters. The colors are set based on the proximity to the orthosteric site (the darker, the closer).

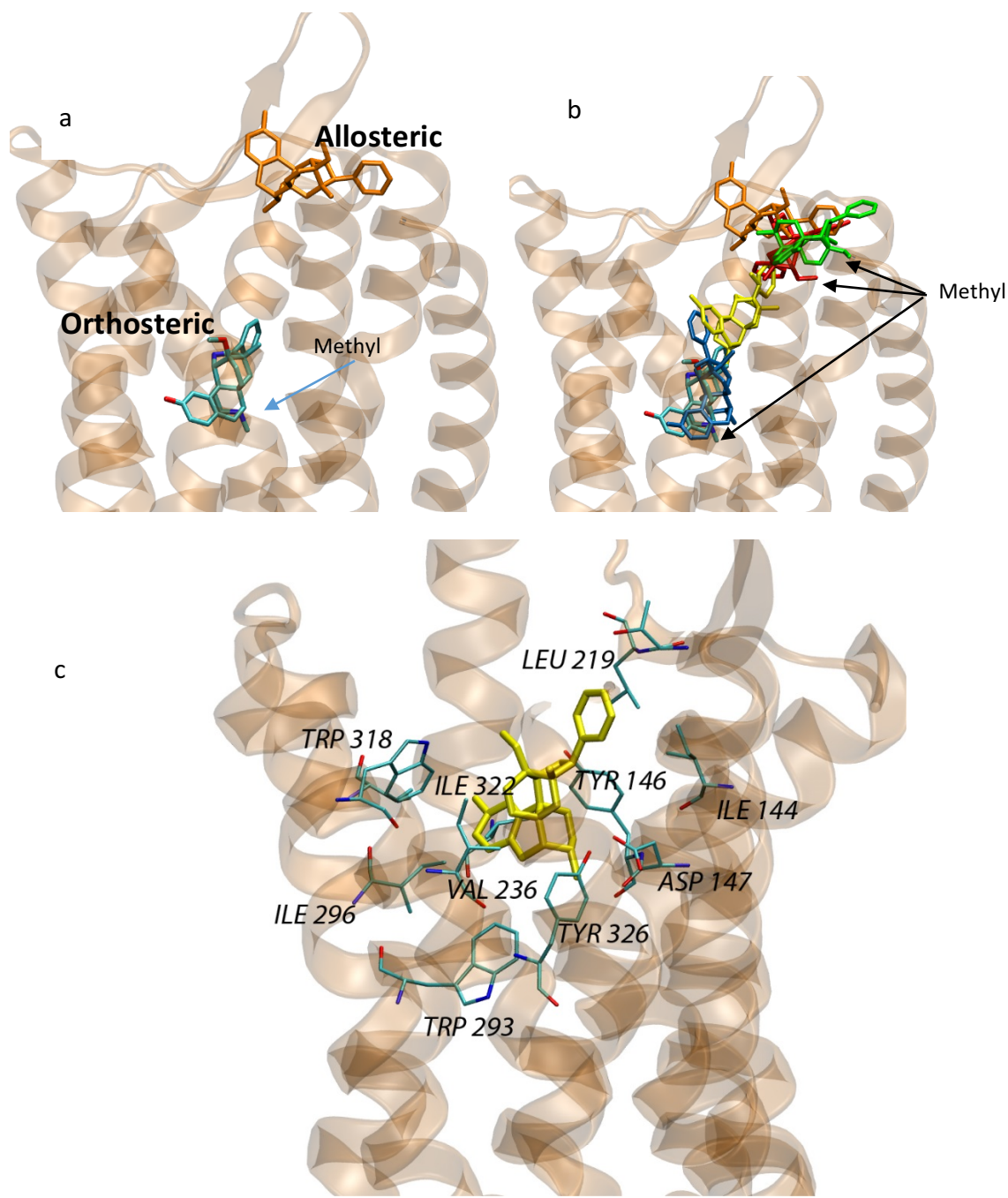

**Figure 5: a|** Respective position and orientation of BU72 in allosteric and orthosteric sites of µOR. **b|** Multiple snapshots of BU72 obtained from MSM clusters on the path from allosteric to orthosteric sites. Each color represents different cluster states (in our MSM model, the clusters are [47, 93, 48, 9, 43] plus the allosteric and orthosteric initial seed poses. **c|** Snapshot of BU72 at the high energy barrier point of its path ((-0.2, 0.8) in Figure 3). The ligand (BU72) is colored in yellow and µOR is depicted in transparent orange. The residues connecting to BU72 at this snapshot is shown and labeled. These residues are important residues extracted from tIC1-tIC7 (see Supporting Information).

# Supporting Information

# for

# Binding Pathway of Opiates to μ-Opioid Receptors Revealed by Unsupervised Machine Learning

*Amir Barati Farimani[§], Evan N. Feinberg[§], Vijay S. Pande[2]*

Department of Chemistry, Stanford University, Stanford, California 94305

[§] The authors equally contributed to this work

1. Simulation set up scheme and ligand chemistry
2. Components of tIC1- tIC4
3. Cluster transition path numbers
4. Distances upon binding from hooking site to binding site

[2] Corresponding Author, e-mail: pande@stanford.edu, web: https://pande.stanford.edu/

## Simulation set up scheme and ligand chemistry

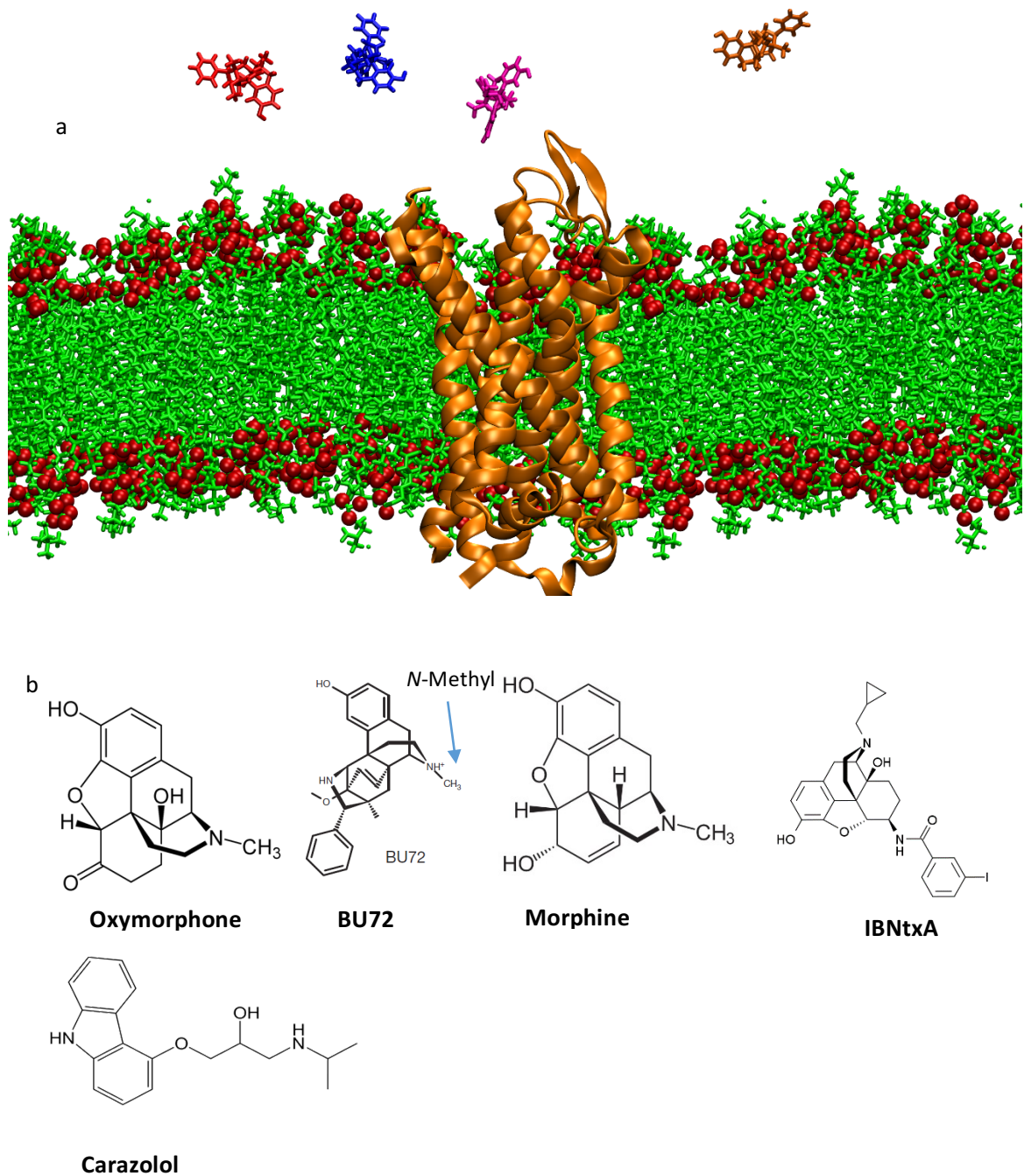

**Figure S.1: a|** Snapshot of simulation (water and ions are not shown) including 4 ligands of the same type (here BU72), lipid, inactive μOR (PDB: 5C1M) **b|** Ligands used in the simulations to find the allosteric binding site (oxymorphone, BU72, morphine, IBNtxA and carazolol).

# Components of tIC1- tIC4

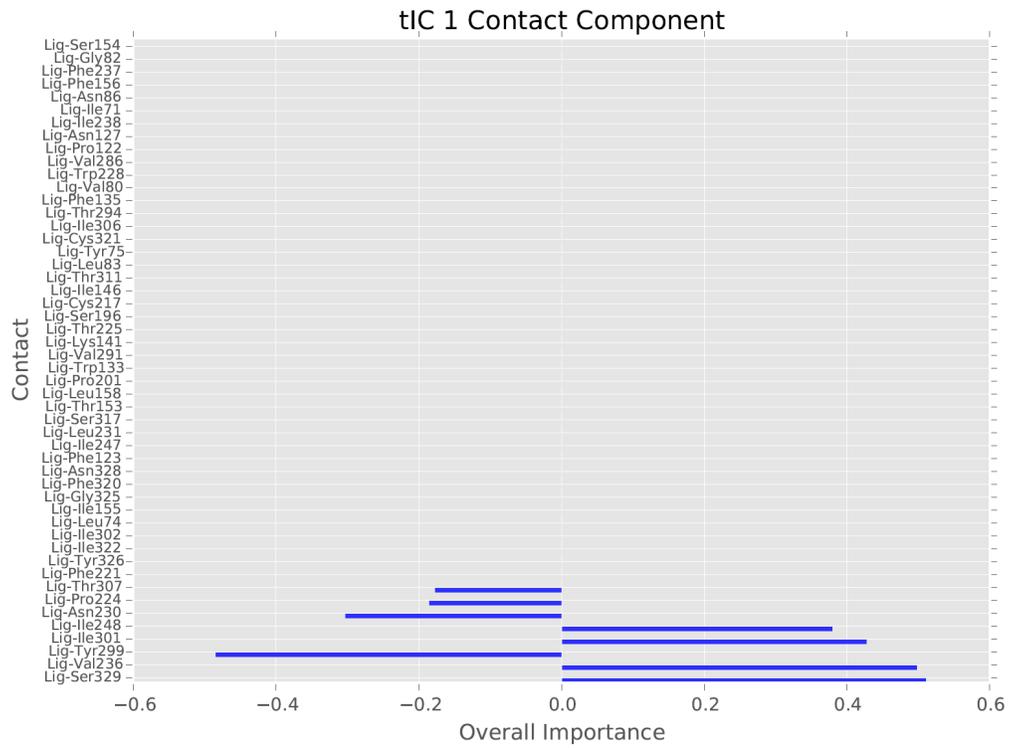

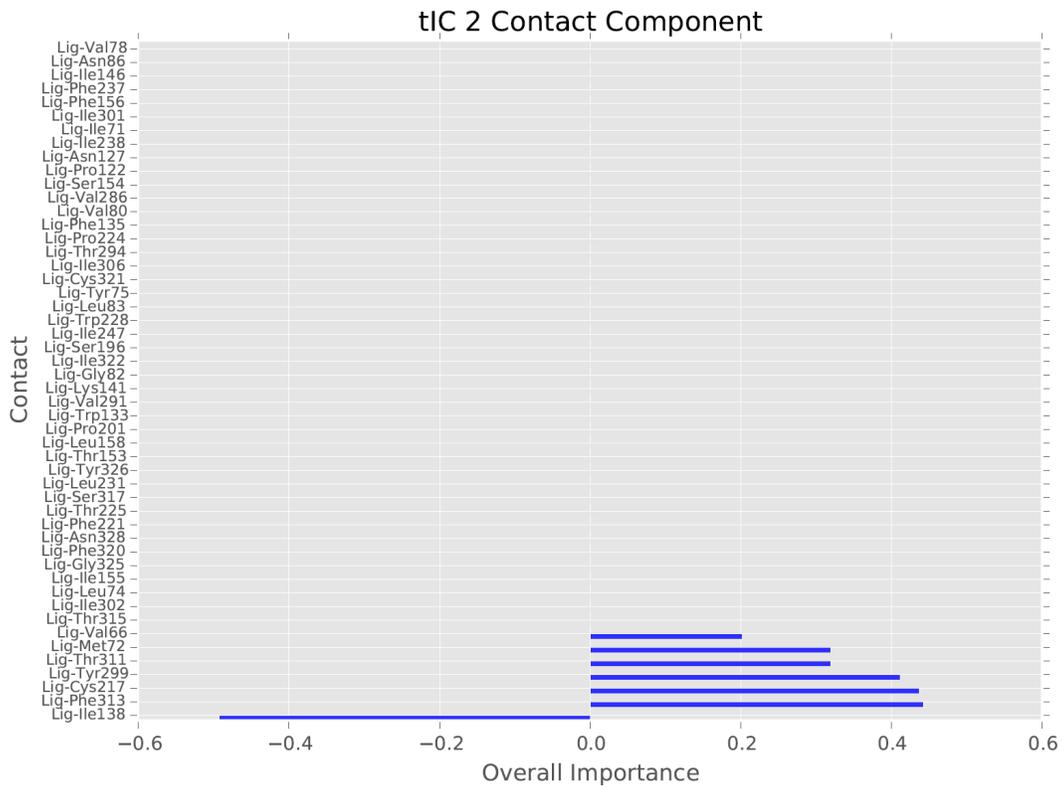

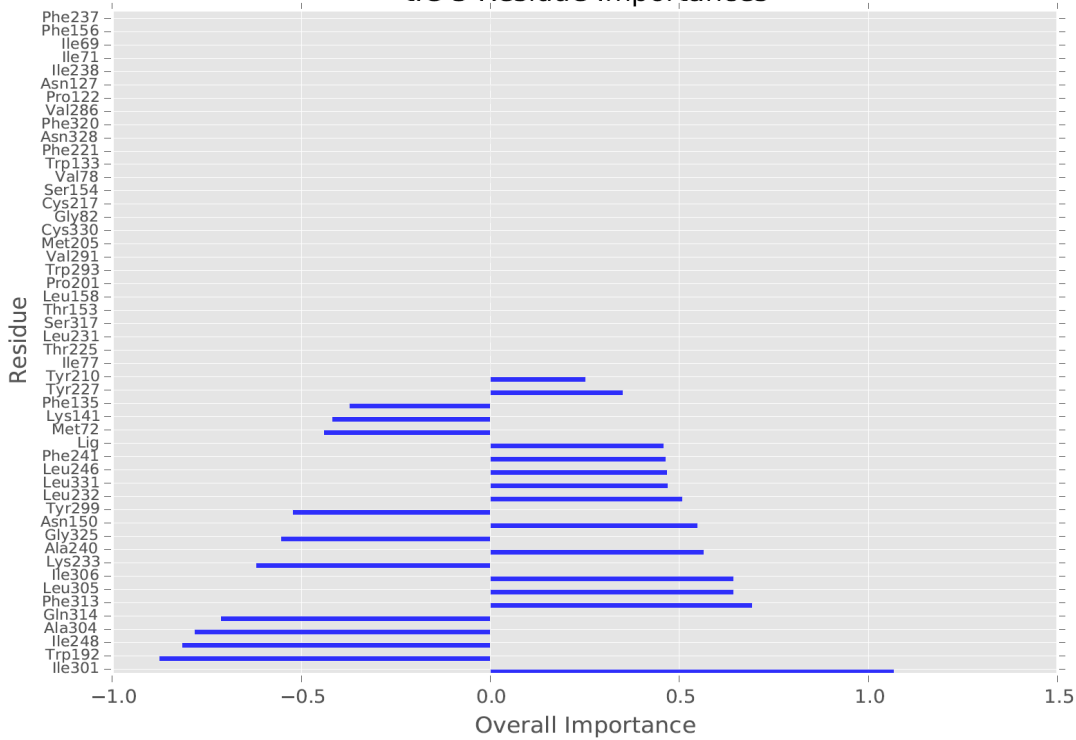
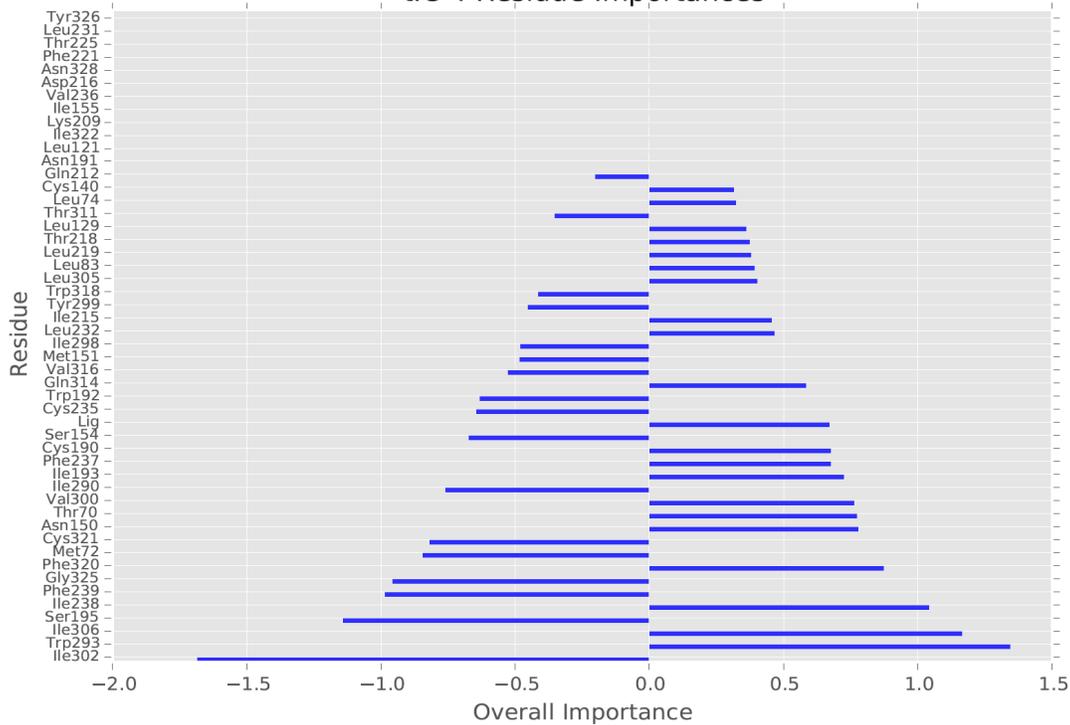

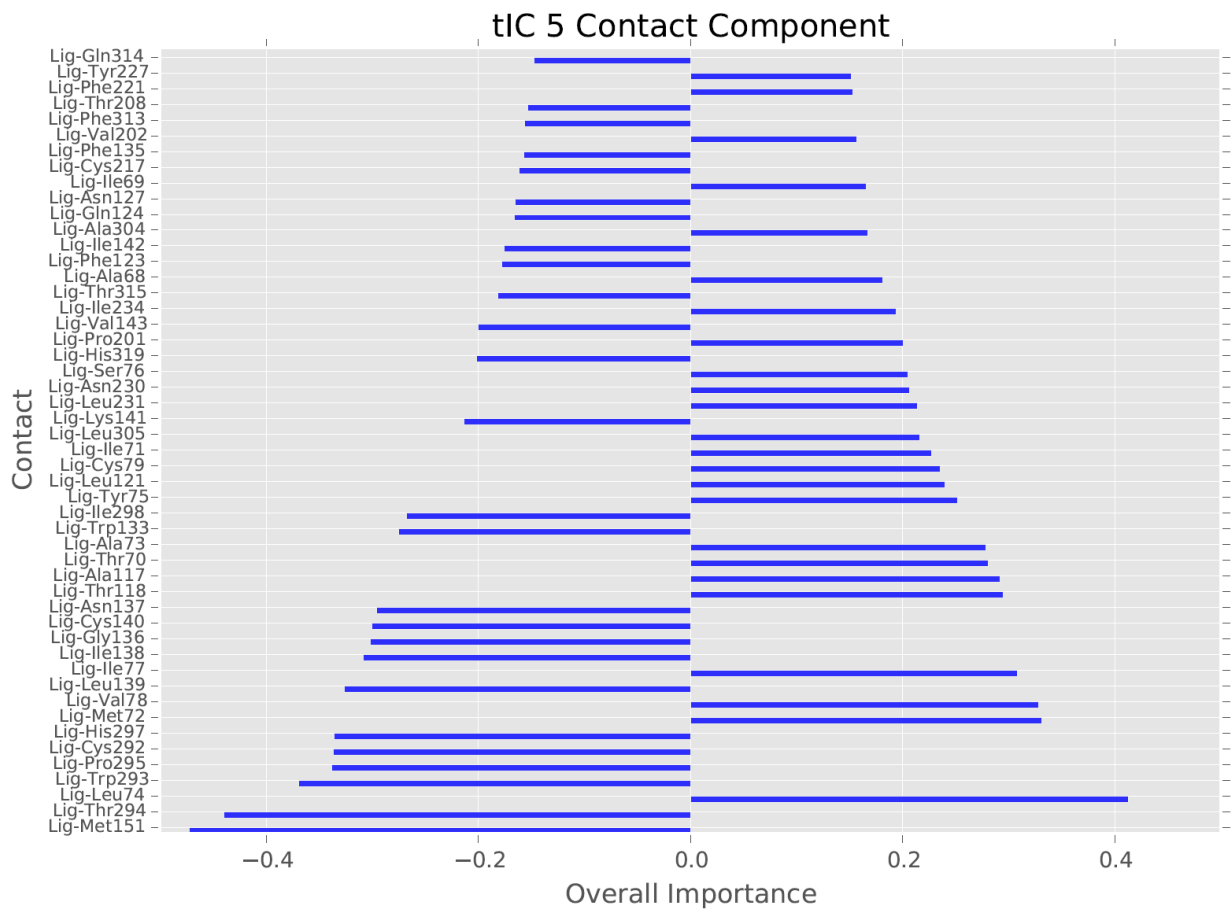

**Figure S.2**. tIC1 to tIC5 representing the important amino acids involved in the Bu72 (Lig) communication to traverse from allosteric to orthosteric

### S.2. Cluster transition path numbers

Below, are the paths derived from Transition Path Theory (TPT)[1] that shows the cluster numbers on different possible paths from MSM shows in Figure S.2. The path that we made the ligand snapshots is shown in bold (**Path8 = [47 93 48 9 43]**)

Path1 = [61 49 73]

Path2 = [61 14 92 23 52 42 43]

Path3 = [61 62 14 92 40 57 42 43]

Path4 = [61 25 78 28  9 43]

Path5= [61 25 49 73]

Path6 = [61 93 96  9 43]

Path7 = [61 62 40 57 42 34]

**Path8 = [47 93 48  9 43]**

Path9 = [47 25 62 92 57 42 21]

Path10 = [72 25 14 40 85 65 57 42 43]

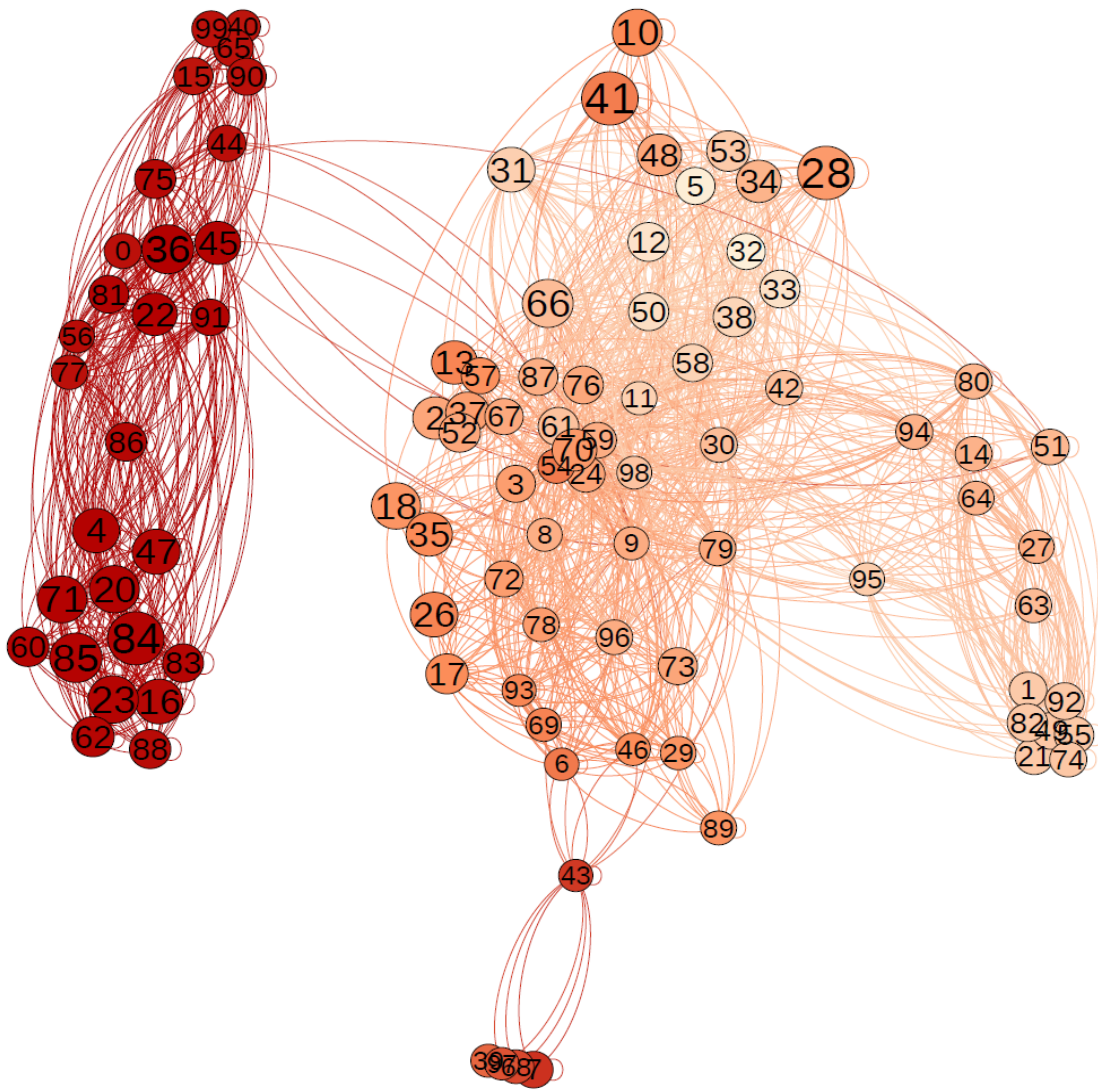

**Figure S.3**. Cluster numbers and the possible paths from allosteric to orthosteric site. The nodes are clusters taken from MSM. Edges show the possible paths.

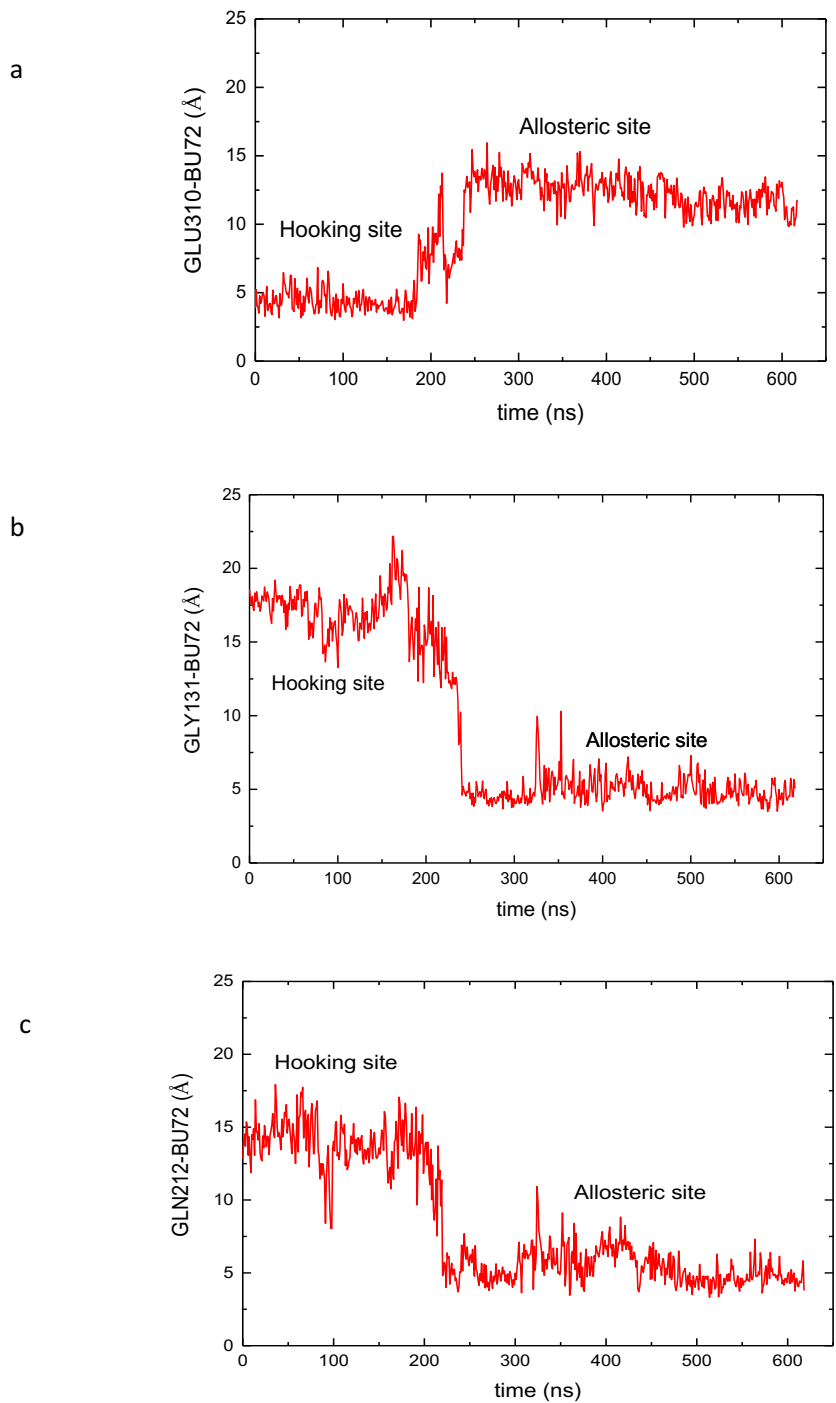

**Figure S4. a**|GLU130-BU72 distance upon binding to hooking site and allosteric sites **b**| GLY131-BU72 distance upon binding to hooking site and allosteric sites. **c**| GLN212-BU72 distance upon binding to hooking site and allosteric sites